\documentstyle[epsf]{article}
\makeatletter
\def\setcaption#1{\def\@captype{#1}}
\makeatother

\begin{document}


\medskip

{\center \Large Study of the atmospheric neutrino flux
                in the multi-GeV energy range\\ }

\bigskip

{\center \large The Super-Kamiokande Collaboration \\}

\bigskip

\begin{center}
Y.Fukuda$^a$, T.Hayakawa$^a$, E.Ichihara$^a$, K.Inoue$^a$,
K.Ishihara$^a$, H.Ishino$^a$, Y.Itow$^a$,
T.Kajita$^a$, J.Kameda$^a$, S.Kasuga$^a$
% ...

\footnotesize \it

$^a$Institute for Cosmic Ray Research, University of Tokyo, Tanashi,
Tokyo 188-8502, Japan\\
% ...

\end{center}

\section*{Abstract}
The flavor ratio of the atmospheric neutrino flux and its
zenith angle dependence have been studied in the multi-GeV energy
range using an exposure of 25.5 kiloton-years of the Super-Kamiokande
detector.  By comparing the data to a detailed Monte Carlo simulation,
the ratio $(\mu/e)_{DATA}/(\mu/e)_{MC}$ was measured to be
$0.66\pm0.06(stat.)\pm0.08(sys.)$.  In addition, a strong distortion in the
shape of the $\mu$-like event zenith angle distribution was observed.
The ratio of the number of upward to downward
$\mu$-like events was found to be
0.52$^{+0.07}_{-0.06}(stat.)\pm0.01(sys.)$, with an expected value of
$0.98\pm0.03(stat.)\pm0.02(sys.)$, while the same ratio for the
$e$-like events was consistent with unity.

\section*{Introduction}
\newcommand{\gsim}{\mathrel{\rlap{\raisebox{.3ex}{$>$}}
    \raisebox{-.6ex}{$\sim$}}}
\newcommand{\lsim}{\mathrel{\rlap{\raisebox{.3ex}{$<$}}
    \raisebox{-.6ex}{$\sim$}}}

Cosmic ray interactions in the atmosphere produce neutrinos. The
flavor ratio of the atmospheric neutrino flux,
$(\nu_{\mu}+\bar{\nu}_{\mu})/(\nu_{e}+\bar{\nu}_{e})$ has been
calculated to an accuracy of better than 5\% in the range from 0.1~GeV
to higher than 10~GeV\cite{ref:hon95,ref:gai96}.  The calculated flux
ratio has a value of about two for energies $\lsim$1~GeV and increases
with increasing neutrino energy.  For neutrino energies higher than a
few GeV, the fluxes of upward and downward going neutrinos are
expected to be nearly equal; geomagnetic field effects on atmospheric
neutrinos in this energy regime are expected to be small because the
primary cosmic rays that produce these neutrinos have rigidities
exceeding the geomagnetic cutoff rigidity ($\sim$10~GeV/$Ze$).

The flavor ratio of the atmospheric neutrino flux has been studied by
several massive underground detectors. In these experiments, the ratio
of the number of $\mu$-like to the number of $e$-like neutrino
interactions observed in the detector was compared with Monte Carlo
(MC) simulation; i.e., the ratio $R\equiv(\mu/e)_{DATA}/(\mu/e)_{MC}$
was measured to study the atmospheric neutrino flavor ratio
$(\nu_{\mu}+\bar{\nu}_{\mu})/(\nu_{e}+\bar{\nu}_{e})$.  In the
measurement of $R$, uncertainties in the neutrino flux and cross
sections cancel.  The expected value for $R$ is unity if there is
agreement between the experiment and the theoretical prediction. Three
experiments -- Kamiokande \cite{ref:kam1}, IMB-3 \cite{ref:imb} and
Soudan 2 \cite{ref:soudan} -- observed an $R$ smaller than unity in
the energy region $E_{\nu}\lsim$ 1 GeV, although two experiments --
Frejus \cite{ref:frejus} and NUSEX \cite{ref:nusex} -- reported no
deviation from unity with smaller data samples.  Super-Kamiokande
\cite{ref:subgev} recently reported a small $R$ for $E_{vis}<$ 1.33 GeV.   
The measured small values of $R$ suggest the possibility of neutrino
oscillations.

The value of $R$ in the ``multi-GeV'' ($E_{\nu}\gsim$ 1 GeV) energy
region has been studied by fewer experiments.  Kamiokande
\cite{ref:kam94} observed a value of $R$ smaller than unity, as well
as a dependence of this ratio on the zenith angle. Since there is a
large difference in the neutrino path-length between upward-going
($\sim$~10,000 km) and downward-going neutrinos ($\sim$ 20 km), a
zenith angle dependence of $R$ can be interpreted as additional
evidence for neutrino oscillations.  IMB-3 has reported a result in a
similar energy range \cite{ref:imb97}, but its smaller data sample
neither confirmed nor ruled out the Kamiokande results.

In this paper, the atmospheric neutrino measurement in the multi-GeV
energy range from the Super-Kamiokande detector is presented.  With an
analyzed fiducial volume exposure of 25.5 kiloton-years, $R$ and the
zenith angle dependence of the flux in this region were measured with
much higher statistics than previous experiments.  In addition to
higher statistics, due to the much larger dimensions of the detector,
Super-Kamiokande can contain multi-GeV muon
events, making possible for the first time a measurement of the momentum
spectrum of $\mu$-like events up to $\sim$8~GeV/$c$.

Super-Kamiokande is a cylindrical 50 kiloton ring imaging water
Cherenkov detector.  The detector consists of an inner detector volume
completely surrounded by an outer detector layer.  The two detectors are
optically separated by a pair of opaque sheets which enclose a dead
region 55~cm in thickness.  The inner detector is 36.2~m high and
33.8~m in diameter; these dimensions are sufficient to contain muons
of momentum up to 8~GeV/$c$.  The inner detector is lined with 11,146
50~cm diameter photomultiplier tubes (PMT).  The photocathode coverage
of the inner wall surface is 40\%.  The outer layer of water is 2.6
$-$ 2.75 m thick and is instrumented with 1885 outward facing 20 cm
diameter PMTs.  To maximize light collection, a reflective surface of
Tyvek covers the walls of the outer detector, and each PMT is fitted
with a 60 cm $\times$ 60 cm plate of wavelength shifter.  The outer
detector is used to reduce background entering from the
surrounding rock and to identify penetrating
muons.  The trigger required at least 29 inner detector PMT hits,
corresponding to the mean number of hit PMTs for a $\sim$~5.7~MeV
electron. The atmospheric neutrino data reduction was applied to $\sim
4 \times 10^8$ raw input triggers from 414 live days of exposure.

\section*{Data reduction}
\begin{figure}
\begin{center}
  \epsfxsize=15cm
  \epsfbox{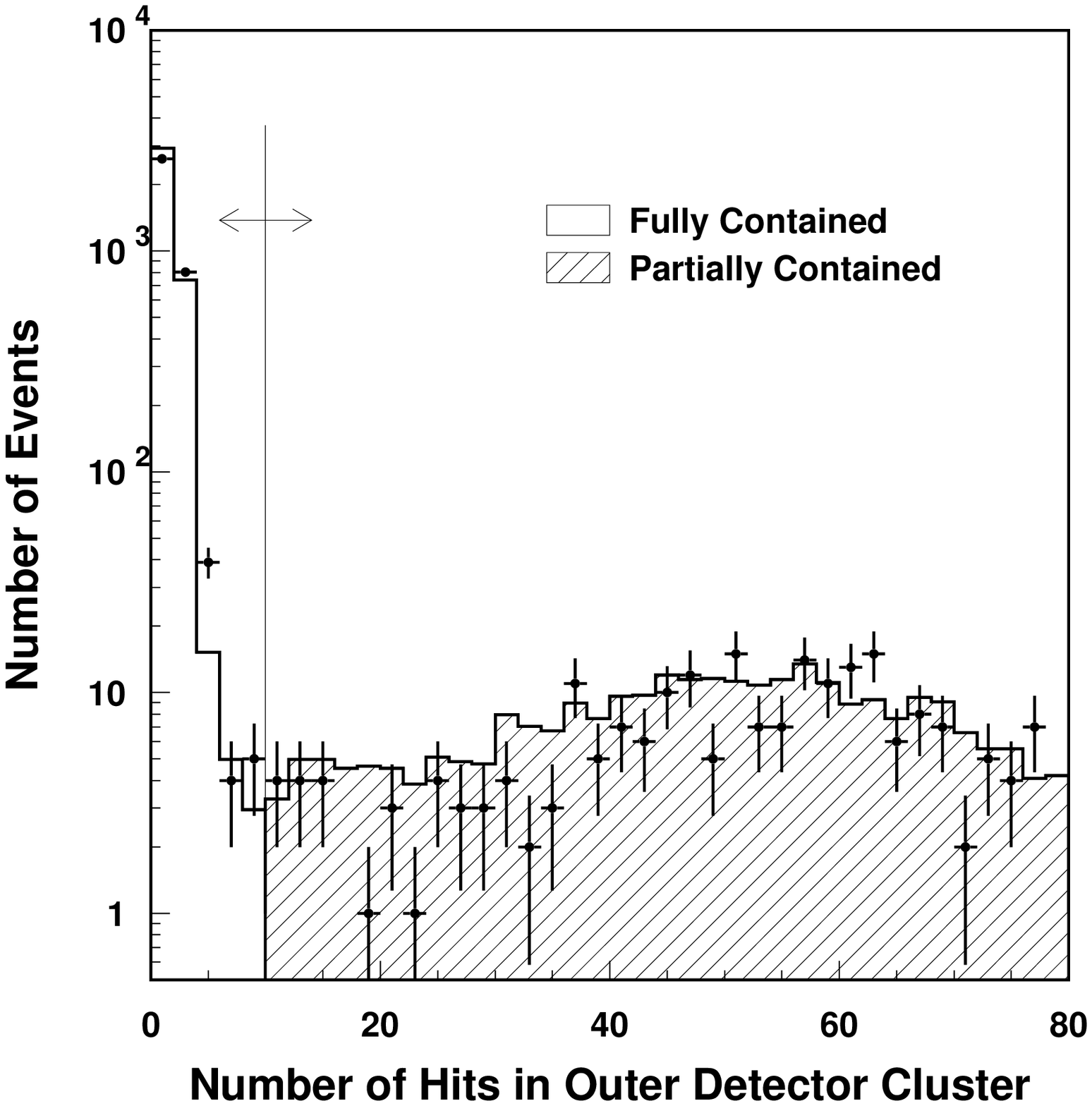}

  \caption{ Distribution of the number of hits within a 500~ns window
in an outer detector cluster for atmospheric neutrino Monte Carlo
(histogram) and data (points).  The Monte Carlo is normalized to the
experimental livetime in this and subsequent plots.
The separation between FC and PC
events is made at 10 hits.}

  \label{odclust} 
\end{center}
\end{figure}

Atmospheric neutrino events have two basic topologies which determine
the data reduction stream.  If all of the visible energy is contained
within the inner detector, the event is called ``fully contained''
(FC).  An event for which some of the produced particles deposit
visible energy in the outer detector is called ``partially contained''
(PC).  More precisely, a clustering algorithm is applied to the hits
in the outer detector: $\geq$~10~hits are required in an outer
detector cluster for an event to be classified as PC, or $<$~10~hits
to be classified as FC.  Figure~\ref{odclust} shows the distribution
of the number of hits in an outer detector cluster for atmospheric
neutrino interactions.  The separation between FC and PC events is
clearly seen.  For both topologies, the interaction vertex is required
to be inside the 22.5~kton fiducial volume, defined as the volume 2~m
from the inner detector PMT planes.

The multi-GeV FC reduction and reconstruction chains are identical to
those used for the sub-GeV event sample, which have been described in
detail in another paper~\cite{ref:subgev}.  The multi-GeV FC data
differed from the sub-GeV FC data only in that we required $E_{vis}$
$>$1.33~GeV.  The total number of multi-GeV FC events in the fiducial
volume was 792.

According to Monte Carlo estimates, the PC data is a 98$\pm$0.3\% pure
sample of charged current (CC) $\nu_{\mu}+\bar{\nu}_{\mu}$ scattering
events. PC events are typically characterized by a single muon with
energy sufficient to escape the inner detector.  The data reduction
for PC events differed significantly from the reduction for FC events,
mainly due to the presence of additional hits in the outer detector.
Because of these extra hits from the exiting muon, a simple criterion
based on the number of hit outer detector tubes could not be used to
reject cosmic ray background. Several automated data reduction
criteria were used to eliminate background in the PC sample before a
final reduction by physicist scanners based on a visual display of the
event data.

(i) Low energy events with fewer than 1000 total p.e. were removed,
corresponding to muons (electrons) with momentum less than 310(110)
MeV/$c$. By definition, an exiting (PC) particle must have reached the
outer detector from the inner fiducial volume, and so must have had a
minimum track length of about 2.5 meters (corresponding to muons with
$\gsim700$ MeV/$c$ momentum).

(ii) The time distribution and spatial clustering of hits in the inner
and outer detectors were used in the next reduction step.  Events for
which the width of the time distribution of hits in the outer detector
exceeded 240 nanoseconds were rejected, as well as events with two or
more spatial clusters of outer detector hits.  These cuts eliminated
many through-going muons, which typically left two well separated
clusters in the outer detector.  Muons which clipped the edges of the
detector were eliminated based upon the topology of the outer detector
cluster.  Cosmic ray muons which entered and stopped in the inner
volume of the detector were eliminated by excluding events with a
relatively small number of inner detector photoelectrons near the
outer detector cluster (1000~p.e. within 2~m).  This cut did not remove
PC neutrino events because PC events produced large numbers of photoelectrons
in the region where the particle exited.

(iii) In the next step, a simple vertex fit and charge weighted
direction estimate were used.  A requirement of $\leq 10$ hits in
the outer detector within 8 meters of the back-extrapolated entrance
point was imposed.  The remaining background after this cut consisted
of muons which left few or no entrance hits in the outer detector.
These events were rejected by requiring the angle subtended by the
earliest inner detector PMT hit, the vertex, and the back-extrapolated
entrance point to be $>37^\circ$.  Remaining corner clipping muons
were rejected by requiring a fitted vertex at least 1.5 meters away
from the corners of the inner detector volume.  A through-going muon
fitter was also applied to reject events with a well fit muon track
greater than 30 meters long.

(iv) A precise automatic fitting algorithm was applied to further
reject entering events, again requiring $\leq$10 hits in the outer
detector within 8 meters of the more accurately back-extrapolated
entrance point. At this stage, a minimum requirement of 3000 total
p.e. was applied.  This requirement corresponded to 350 MeV of visible
energy, well below that of any exiting muon.  It was estimated that
0.1\% of the PC events in the fiducial volume were eliminated by this
requirement. After this step, 758 events remained in the full detector
volume.

(v) The remaining events were scanned with an interactive graphical
event display to eliminate the remaining background.  There were two
independent scans of the data, with Monte Carlo events interspersed
randomly and in proportion to livetime.  Scanners were asked to
classify each event as a neutrino (FC or PC), or various types of
known background.  Most ($>$85\%) of the events eliminated by scanning
were entering events (through-going or stopping muons).  A third and
final scan was used to resolve disputes between the first two
independent scans.  Blind scanning of Monte Carlo events mixed in with
the data showed that the scanning efficiency was $\ge$99\% in the
fiducial volume.

Applying the same reduction steps to Monte Carlo generated
atmospheric neutrino events yielded an 
overall data reduction efficiency of 88$\pm$5\% for
interactions in the fiducial volume.  Inefficiency accrued at the few
percent level in each automated reduction stage.  After the visual
scan, 352 (230) PC events remained in the full (fiducial) detector
volume.

The vertex resolution for the FC single-ring $\mu$-like ($e$-like) events
was estimated to be 23 (42) cm and for PC events it was estimated to be
104 cm.

\section*{Flavor ratio}
For the $\nu_{\mu}/\nu_{e}$ ratio in the multi-GeV region, we used
FC single-ring events with E$_{vis}$ $>$ 1.33 GeV and
PC events. In Table \ref{tb:sum}, the numbers of
observed events are summarized along with the corresponding Monte
Carlo predictions.  For FC events, lepton flavor was determined by the
particle identification method described in Ref.\cite{ref:subgev}.
According to Table \ref{tb:sum}, the PC events comprise a
98\% pure charged-current (CC) $\nu_{\mu}$ sample; therefore we
classify all PC events as $\mu$-like. FC $\mu$-like ($e$-like) events
were a 99\% (84\%) pure CC $\nu_\mu$ (CC $\nu_e$) sample.  The
relatively low purity of the $e$-like sample was due to the
incompleteness of the separation of the electromagnetic shower events
from CC and neutral-current (NC) interactions which produced single and
multiple pions.

From these data, the ratio 
$R_{FC+PC}\equiv(\mu/e)_{DATA}/(\mu/e)_{MC}$ 
 for the multi-GeV region was obtained:
  \begin{eqnarray*}
      R  = 0.66 \pm 0.06(stat.) \pm 0.08(sys.),
 \end{eqnarray*}
where $e$ is the number of FC $e$-like events and $\mu$ is the 
sum of the numbers of FC and PC $\mu$-like events.
The same ratio for FC events only is $R_{FC}$ = 0.64 $\pm$ 0.07 $\pm$
0.10.  This result is consistent with the previous results
from Kamiokande \cite{ref:kam94} in the same energy range and is also
consistent with the results in the lower energy region obtained by Kamiokande \cite{ref:kam1}
IMB-3 \cite{ref:imb}, Soudan-2 \cite{ref:soudan} and Super-Kamiokande
\cite{ref:subgev}.  It is significantly smaller than unity.

\newcommand{\nue}{\mbox{${\nu}_{e}$}}
\newcommand{\neb}{\mbox{$\overline{\nu}_{e}$}}
\newcommand{\num}{\mbox{${\nu}_{\mu}$}}
\newcommand{\nmb}{\mbox{$\overline{\nu}_{\mu}$}}

The systematic uncertainty in $R_{FC+PC}$ came from several sources: 
5\% from uncertainty in the atmospheric flux ratio $(\num+\nmb)/(\nue+\neb)$,
4.3\% from the
uncertainties in the CC neutrino cross section and
nuclear effects in H$_{2}$O targets,
4.1\% from
corresponding uncertainties for NC interactions,
1.6\% from the uncertainty of the cosmic ray primary energy spectrum,
1\% from hadron tracking simulation,
6\% from the single and multi-ring separation for FC events, 
4\% from energy scale determination
and 0.5\% from energy resolution uncertainty,
3\% from PC reduction efficiency,
3\% from particle type misidentification,
2\% from the vertex reconstruction,
1\% from uncertainty in the separation between FC and 
PC events,
and 3\% from Monte Carlo statistical uncertainty.
In addition, uncertainties on $R$ from event sample
contamination included: 1\% from cosmic muons,
less than 0.1\% from the neutron-induced background, and less than 0.5\% 
from ``flasher'' events (Ref.\cite{ref:subgev}).
Adding all of these uncertainties in quadrature,
we estimated the total systematic uncertainty on $R$ to be 12\%.

\begin{figure}
\begin{center}
  \epsfxsize=15cm
  \epsfbox{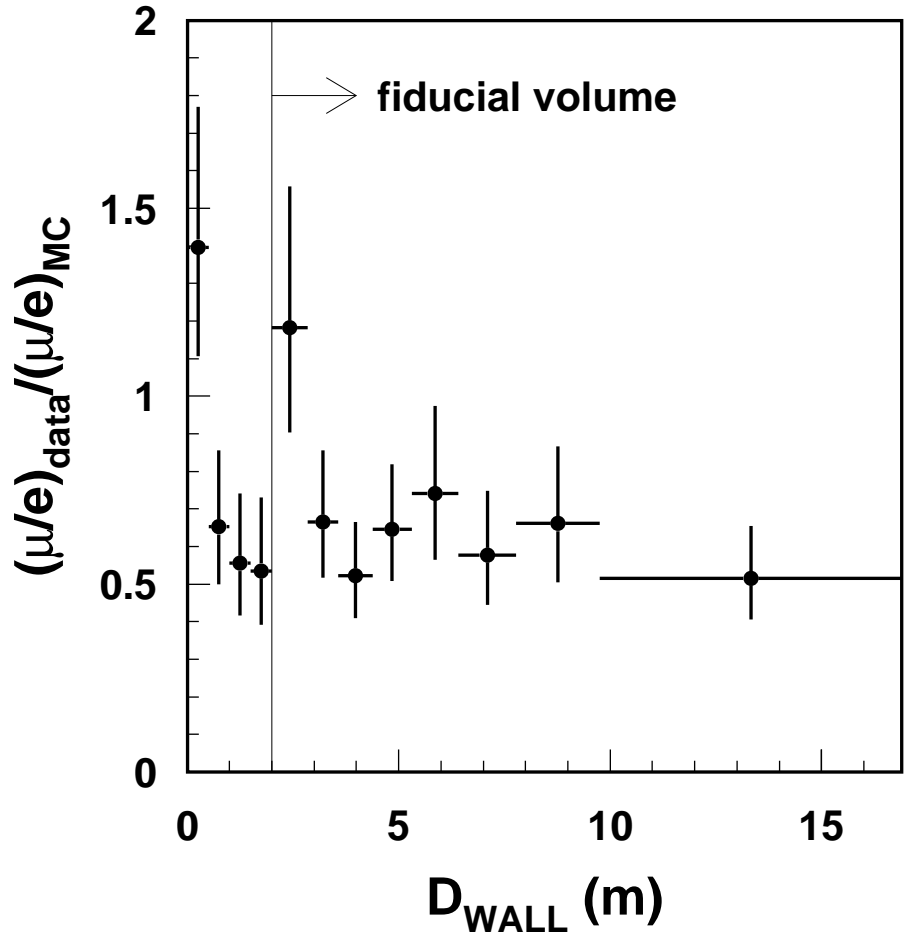}
  \caption{
    ${R_{FC+PC}}$ as a function of $D_{WALL}$, the distance between
    the event vertex and the nearest inner detector wall. The region
    $D_{WALL}~>$ 2~m is the fiducial volume. Error bars show the
  statistical uncertainties of the data and Monte Carlo.}
  \label{dwall} 
\end{center}
\end{figure}

Fig. \ref{dwall} shows $R_{FC+PC}$ as a function of the distance of the
reconstructed vertex to the nearest inner detector wall $D_{WALL}$.
From this figure,
there is no evidence for neutron, gamma-ray, or cosmic ray muon
background which could change $R$ near the edges of the fiducial
volume. By scanning FC and PC events, it was determined that the high
$R$ value of the bin nearest to $D_{WALL}$=0 was mostly due to a
contamination of cosmic ray muon background which entered into the
inner detector through less efficient regions of the outer detector.
However, the high $R$ value (by 2 standard deviations) at the bin just
inside the fiducial volume was due to a relative deficit of $e$-like
events by 2.0 sigma (34\% deficit) and a 1.1 sigma excess
of FC $\mu$-like and PC events (16\% excess).  Scanning of events in this bin
showed no evidence for background contamination.

\begin{figure}
\begin{center}
  \epsfxsize=10cm
  \epsfbox{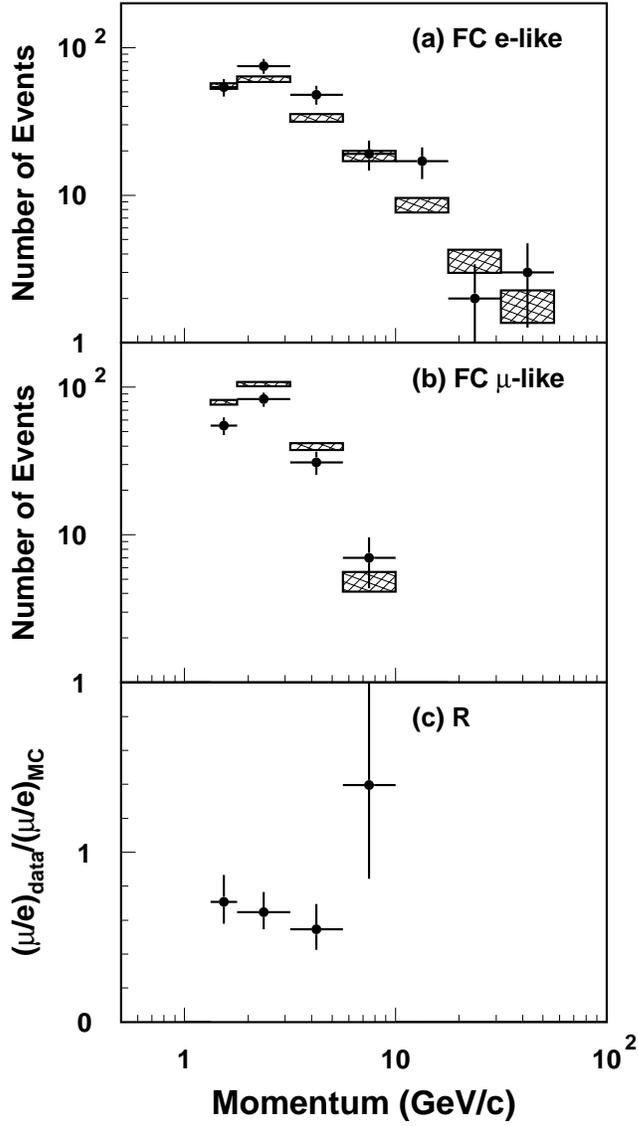}

  \caption{
    Reconstructed momentum distributions for :
    (a) FC $e$-like events, (b) FC $\mu$-like events and 
    (c) ${R_{FC}}$ as a function of momentum. 
    The histograms with shaded error bars show the Monte Carlo predictions with 
    their statistical uncertainties. }
  \label{Rvse}
\end{center}
\end{figure}

\begin{figure}
\begin{center} 
  \epsfxsize=15cm
  \epsfbox{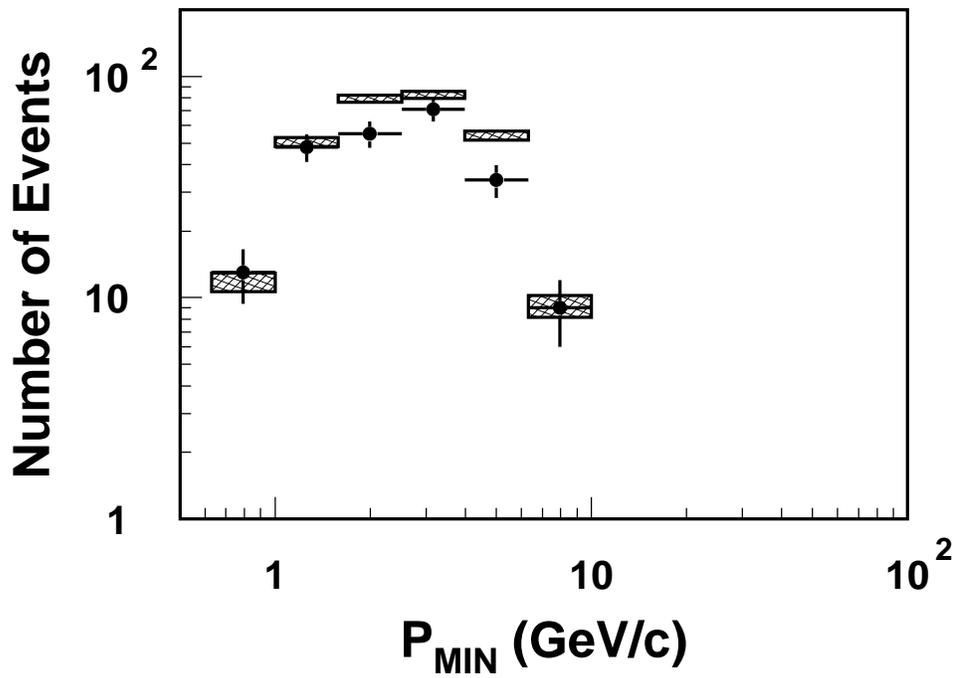}

  \caption{
    Minimum momentum ($P_{MIN}$) distribution for PC events.  
    $P_{MIN}$ is estimated from the reconstructed track length
    from the vertex to the outer detector.
    The histogram with shaded error bars shows the Monte Carlo prediction with 
    its statistical uncertainties.}
  \label{evis}
\end{center} 
\end{figure}

For FC events, lepton energy was reconstructed with 4\% uncertainty
(3\% uncertainty from the estimated energy resolution and 2.5\%
uncertainty from the absolute energy calibration).  Fig. \ref{Rvse}
shows reconstructed momentum distributions for : (a) FC $e$-like
events, (b) FC $\mu$-like events and (c) $R_{FC}$ as a function of
momentum.  The values of $\chi^2$/d.o.f. for comparison of the shapes
of the MC and data distributions were 4.4/6 and 1.4/3 for the $e$-like
and $\mu$-like distributions respectively.  (c) was consistent with a
flat distribution within the statistical uncertainty ($\chi^2$/d.o.f. 
was about 2.3/3). For the PC events, only a minimum energy for
the muon is measurable.  Fig. \ref{evis} shows the minimum momentum
($P_{MIN}$) distribution assuming that the exiting particle is a muon.
The shape is consistent with the Monte Carlo prediction
($\chi^2$/d.o.f. = 3.7/5).  The mean neutrino energies of our sample
were approximately 5 GeV, 3 GeV, 15 GeV and 9 GeV for FC $e$-like, FC
$\mu$-like, PC, and FC $\mu$-like+PC respectively.

\begin{table}[ht]
 \renewcommand{\arraystretch}{1.5}
 \newcommand{\lw}[1]{\smash{\lower2.ex\hbox{#1}}}
 \begin{center}
  \begin{tabular}{lrcrrrr} \hline\hline
     & \lw{Data}\hfil &&\multicolumn{4}{c}{Monte Carlo} \\
     \cline{4-7}
     & & &total    &$\nu_{e}$  CC(q.e.) & $\nu_{\mu}$  CC(q.e.) & NC\\
     \hline
         FC events \\
     \hline
     single ring   &  394 && 411.6 & 155.4(70.1)& 239.7(125.6)&16.6 \\
     {} $e$-like     &   218 &&  182.7 & 154.1(69.9)& 12.9(1.7)&15.6 \\
     {} $\mu$-like &   176 &&  229.0 & 1.2(0.2)& 226.8(123.9)& 0.9 \\
     multi ring    &   398 &&  433.7 & 129.2(9.0)& 237.2(8.6)& 67.1 \\
     \hline
     total         &  792 && 845.2 & 284.5(79.0)&477.0(134.3)& 83.7 \\
     \hline
         \\
        PC events \\
     \hline
     total         &  230 && 287.7 & 4.4(0.8)&281.5(51.6)& 1.9 \\
     \hline\hline
  \end{tabular}
 \end{center}
\caption{Summary of the multi-GeV event sample compared with the
Monte Carlo estimation for 25.5 kt$\cdot$yrs of detector exposure using the
calculated flux from~Ref.\cite{ref:hon95}. Monte Carlo statistics
have been normalized to the live time of the experimental data. ``q.e.''
refers to quasi-elastic events.}
 \label{tb:sum}
 \end{table}

A second FC event analysis using a completely independent analysis
chain was also performed.  The reduction and reconstruction steps were
identical to those performed for the sub-GeV ``Analysis B'' FC data
sample described in Ref.\cite{ref:subgev}.  For Analysis B, which did
not include a PC event selection, the multi-GeV sample comprised
events with $1.33$~GeV$<E_{vis}<5$~GeV.  The total number of multi-GeV
FC events in the fiducial volume was 602.  Among the FC multi-GeV
single-ring events, 100 events were classified as $e$-like and 109
events were classified as $\mu$-like.  Based on an independent Monte
Carlo sample \cite{ref:subgev} of 10.2 years of equivalent exposure,
the predicted numbers of FC $e$-like and $\mu$-like events were 91.6
and 141.1, respectively.  The independent analysis result was
$R_{FC}=0.71^{+0.11}_{-0.09}\pm0.07$.

Differences between the two analyses were consistent with estimated
efficiencies and resolutions.  Although approximately the same live time was
analyzed, there were fewer events in the final Analysis~B sample due
to the upper energy cut and a tighter single-ring selection.  The
Analysis~B result confirmed that the value of $R$ was smaller than unity.

\section*{Zenith angle dependence}
The angular correlation between the neutrino direction and the produced
charged lepton direction for multi-GeV neutrinos is $15-20^\circ$
(RMS). Therefore the zenith angle distribution of the leptons reflects
that of the neutrinos.

\begin{figure}
\begin{center} 
  \epsfxsize=15cm
  \epsfbox{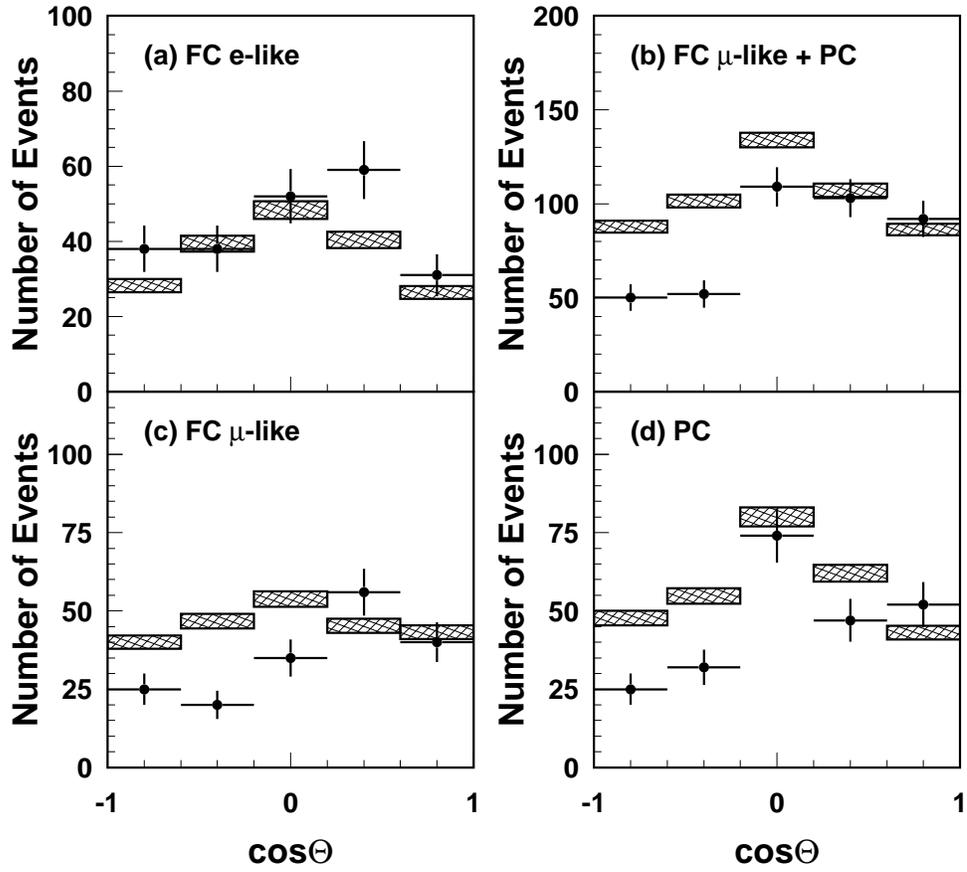}
  \caption{
    Zenith angle distributions for: (a) FC e-like events, 
    (b) FC $\mu$-like and PC events,
    (c) FC $\mu$-like events and (d) PC events. 
    ${\rm \cos\Theta = 1}$ means down-going.
     The histograms with the shaded error bars show the Monte Carlo 
     predictions 
     with their statistical uncertainties.}
  \label{zen}
\end{center} 
\end{figure}

\begin{figure}
\begin{center} 
  \epsfxsize=15cm
  \epsfbox{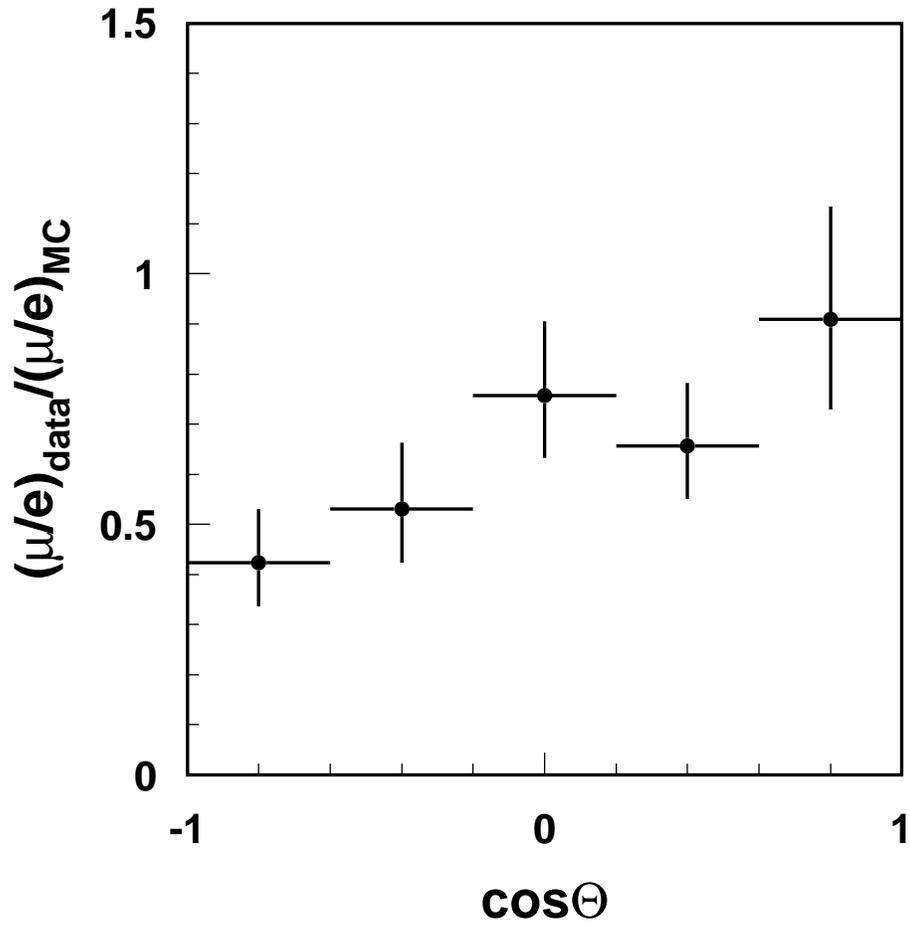}

  \caption{
    Zenith angle dependence of ${\rm R_{FC+PC}}$. 
    Error bars show the statistical uncertainties of the data and Monte Carlo.}
  \label{zenr}
\end{center} 
\end{figure}

Fig. \ref{zen} shows the $\cos\Theta$ distribution for (a) FC $e$-like
events, (b) FC $\mu$-like +PC events, (c) FC $\mu$-like events and (d)
PC events, where $\Theta$ is the zenith angle of the particle direction,
and $\cos\Theta = -1(+1)$ corresponds to upward-going (downward-going).  The
neutrino flux calculation has a $\sim$20\% uncertainty in absolute flux;
therefore we compared only the shape of the zenith angle distribution for data
and Monte Carlo.  The $\chi^2$ values for the shape analysis were : 
$\chi^2$/d.o.f.
= 4.5/4, 27.3/4, 20.0/4 and 16.0/4, for Figs.~\ref{zen} (a) through (d)
respectively.  For $e$-like events, the data were consistent with Monte Carlo,
but for $\mu$-like events, there was an obvious discrepancy. Fig.
\ref{zenr} shows $R$ as a function of zenith angle. 
The value of $\chi^2$/d.o.f. with respect to a flat line through
the mean is about 8.2/4.
Fig. \ref{zen} and
Fig. \ref{zenr} show that there is a significant deficit in
the number of upward-going $\mu$-like
events relative to the number of downward-going ones. 
The ratio of the number $N_{up}$ of upward-going
($-1<\cos\Theta<-0.2$) events to the number $N_{down}$ of downward-going
($0.2<\cos\Theta<1$) events is shown in Table~\ref{updownratio}.
Horizontal-going events ($-0.2<\cos\Theta<0.2$) were excluded.  A significant
deficit was present in $N_{up}/N_{down}$ for
both FC $\mu$-like and PC events.  The value of
$N_{up}/N_{down}$ for $e$-like events was consistent with expectations.  The
statistical significance of the asymmetry for $\mu$-like data
was 5.7$\sigma$.  These results confirmed the Kamiokande results
\cite{ref:kam94} with smaller statistical uncertainty.\footnote{From
Fig. 3 of Ref. \cite{ref:kam94}, the Kamiokande $N_{up}/N_{down}$ value for the
multi-GeV $\mu$-like ($e$-like) data was 0.58$^{+0.13}_{-0.11}$
(1.38$^{+0.39}_{-0.30}$).}

Several sources of systematic uncertainty on the up/down ratio were
considered.  First, the up/down ratios for $e$-like events, $\mu$-like
events and $(\mu / e)_{MC}$ for the two calculated
fluxes~\cite{ref:hon95,ref:gai96} were compared. Both calculations
predicted up/down ratios very close to unity.  The predicted
$N_{up}/N_{down}$ ratio differed between the two calculations by 2\%
and $<$1\% for the $e$-like and $\mu$-like events, respectively, and
the up/down ratio of $(\mu /e)_{MC}$ differed by $<$2\% between the
two calculations.  These two calculations do not assume the existence
of a 1 km mountain over the Super-Kamiokande detector; the rock
reduces the neutrino flux due to muons which are stopped before they
can decay in flight.  We estimated the effect of the presence of rock
on the predicted flux: $N_{up}/N_{down}$ was increased by about 2\%
and 1.5\% for $e$-like and $\mu$-like events respectively.  The
presence of rock changed the up/down ratio of $(\mu /e)_{MC}$ by less
than 1\%.  We did not expect any other significant sources of
systematic uncertainty in the predicted up/down ratio.  The total up/down
systematic uncertainty in the Monte Carlo is shown in
Table~\ref{updownratio}.

We estimated that the detector PMT gain was 3\% higher for down-going
particles than for up-going particles by studying decay electrons from
stopping cosmic ray muons~\cite{ref:subgev}.  This gain asymmetry
caused $\pm$~2.2\% and $\pm$~3.5\% uncertainty in the up/down ratio
for the $e$-like and
FC $\mu$-like events, respectively.  However, the gain asymmetry
caused less than 0.1\% uncertainty in the up/down ratio for the PC events
due to the looseness of the energy cut in the selection of the PC
events. The gain asymmetry caused a 0.6\% uncertainty on 
the up/down ratio of $(\mu
/e)_{DATA}$.  A contamination of non-neutrino background such
as down-going cosmic muons could have
directional correlation.  The maximum contribution to the
uncertainty in the up/down ratio from contamination
was estimated to be $\pm$ 0.5\%,
$\pm$ 2.0\% and $\pm$ 2.1\% for the $e$-like
events, $\mu$-like events, and $(\mu /e)_{DATA}$, respectively.  From
these studies, the total systematic uncertainties in the up/down ratios for
the data were estimated and are shown in Table~\ref{updownratio}. The
total systematic uncertainty in the up/down ratio of 
$R_{FC+PC}=(R_{FC+PC})_{up}/(R_{FC+PC})_{down}$ 
was 3\%,
which was much smaller than the statistical uncertainty.

The validity of the present analysis can be tested by measuring the
azimuth angle distribution of the incoming neutrinos, which is
insensitive to a possible influence from neutrino oscillations.  The
shape of the azimuth angle distributions agreed with the Monte Carlo
predictions which were nearly flat.  The shape comparison
$\chi^2$/d.o.f. values were 9.1/7, 3.3/7 and 3.7/7 for $e$-like, FC
$\mu$-like and PC events, respectively. Therefore, the only observed
directional distortion was for the $\mu$-like zenith angle
distribution.

\begin{table*}
\renewcommand{\arraystretch}{1.5}
\begin{center}
\begin{tabular}{cllrrrr}
\hline
\hline
\multicolumn{1}{c}{ } & \multicolumn{1}{c}{ } & \multicolumn{1}{c}{ }
& \multicolumn{1}{c}{$N_{up}$} & \multicolumn{1}{c}{$N_{down}$} &
\multicolumn{1}{c}{$N_{up}/N_{down}$} 
& \multicolumn{1}{c}{$\frac{N_{up}-N_{down}}{N_{up}+N_{down}}$} \\ 
\hline
     &e-like      &         &  76 & 90 & 0.84 $^{+0.14}_{-0.12}$ $\pm$0.02 & $-$0.084 $\pm$ 0.077 $\pm$ 0.01 \\
\cline{2-7}
data &            & (FC+PC) &   102 &   195 & 0.52 $^{+0.07}_{-0.06}$ $\pm$0.01 & $-$0.313 $\pm$ 0.055 $\pm$ 0.01\\
     & $\mu$-like & (FC)    &    45 &   96 & 0.47 $^{+0.09}_{-0.08}$ $\pm$0.02 & $-$0.362 $\pm$ 0.079 $\pm$ 0.02\\
     &            & (PC)    &    57 &   99 & 0.58 $^{+0.10}_{-0.09}$ $\pm$0.01 & $-$0.269 $\pm$ 0.077 $\pm$ 0.01\\
\hline
     & e-like     &         &  67.6 &  66.8 & 1.01 $\pm$0.06 $\pm$0.03 & 0.006 $\pm$ 0.029 $\pm$ 0.01\\
\cline{2-7}
Monte Carlo   &            & (FC+PC) & 189.3 & 193.6 & 0.98 $\pm$0.03 $\pm$0.02 & $-$0.011  $\pm$ 0.017 $\pm$ 0.01\\
     & $\mu$-like & (FC)    &  86.8 &  88.5 & 0.98 $\pm$0.05 $\pm$0.02 & $-$0.010  $\pm$ 0.025 $\pm$ 0.01\\
     &            & (PC)    & 102.5 & 105.1 & 0.98 $\pm$0.05 $\pm$0.02 & $-$0.013  $\pm$ 0.023 $\pm$ 0.01\\
\hline
\hline
\end{tabular}
\end{center}
\caption{Summary of upward-going and downward-going
events. Upward-going (downward-going) events are those with zenith
angle $-1<\cos\Theta<-0.2$ ($0.2<\cos\Theta<1$).  The $N_{up}/N_{down}$
ratios are shown in the third column with their statistical and
systematic uncertainties.  Also shown in the last column are the
up-down asymmetry $(N_{up}-N_{down})/(N_{up}+N_{down})$ values with
their statistical and systematic uncertainties.}
\label{updownratio}
\end{table*}

For Analysis~B, the up/down ratio for $e$-like events was
1.17$^{+0.34}_{-0.27}\pm$0.01 for data and 0.94$\pm$0.08 for Monte Carlo; the
up/down ratio for $\mu$-like events was 0.38$^{+0.11}_{-0.09}\pm$0.01
for data and 0.99$\pm$0.06 for Monte Carlo.  The Analysis~B result confirmed
the significant up/down asymmetry for $\mu$-like events.

Several studies were undertaken to evaluate whether it is possible to
reproduce the observed distortion of the FC multi-GeV $\mu$-like
zenith angle distribution (in the absence of neutrino oscillations) by
assuming various kinds of possible angular-dependent systematic biases.
Angular-dependent muon detection efficiency, energy and track
reconstruction were considered; in all cases investigated, consistency
between Monte Carlo and data for $\mu$-like events could only be
attained by assuming unrealistically large systematic errors.  From
these studies, we conclude that an angular dependent systematic effect
is unlikely to be responsible for
the distortion of the $\mu$-like zenith angle distribution.

\section*{Conclusions}
The atmospheric neutrino data in the multi-GeV energy range collected
from the Super-Kamiokande detector for the first 414 livetime-days are
presented in this paper. The mean value of $R$ was significantly
smaller than unity, $R_{FC+PC} = 0.66 \pm 0.06 (stat.) \pm 0.08
(sys.)$.  In addition, a strong deviation from expectation in the
shape of the $\mu$-like event zenith angle distribution was observed.
The observed up/down asymmetry of the $\mu$-like events,
$N_{up}/N_{down}$ = 0.52 $^{+0.07}_{-0.06}\pm$ 0.01, deviated from an
expected up/down symmetry, whereas the $e$-like distribution was
consistent with the expected up/down symmetry.  Two independent
analyses yielded consistent results on these effects.  While the
zenith angle dependence of the $\mu$-like data cannot be explained by
any plausible systematic detector effect considered, the relative
deficit of upward-going $\mu$-like events from neutrinos that traveled
a long distance suggests the disappearance of $\nu_{\mu}$ via neutrino
oscillations.

\vspace{0.1in}
We gratefully acknowledge the cooperation of the Kamioka Mining and
Smelting Company. The Super-Kamiokande experiment has been built and
operated from funding by the Japanese Ministry of Education, Science,
Sports and Culture, and the United States Department of Energy.

\unitlength=1mm

\onecolumn
\end{document}